# Auto-balanced common shock claim models

Greg Taylor

School of Risk and Actuarial Studies
University of New South Wales
UNSW Sydney, NSW 2052
AUSTRALIA

Phuong Anh Vu

Taylor Fry
45 Clarence Street
Sydney NSW 2000
AUSTRALIA

**Abstract.** The paper is concerned with common shock models of claim triangles. These are usually constructed as a linear combinations of shock components and idiosyncratic components. Previous literature has discussed the unbalanced property of such models, whereby the shocks may over- or under-contribute to some observations. The literature has also introduced corrections for this. The present paper discusses "auto-balanced" models, in which all shock and idiosyncratic components contribute to observations such that their proportionate contributions are constant from one observation to another. The conditions for auto-balance are found to be simple and applicable to a wide range of model structures. Numerical illustrations are given.

# 1. Introduction

Common shock models were introduced to the actuarial literature by Lindskog and McNeil (2003). The concept has been used in models of claim triangles by Meyers (2007) and Shi, Basu and Meyers (2012). It has also been used by Furman and Landsman (2010) in capital modelling, and by Alai, Landsman and Sherris (2013, 2016) in mortality modelling. Avanzi, Taylor and Wong (2018) discussed the application of common shocks to claim arrays in generality, e.g. shocks with respect to accident, development and payment periods, and others.

Avanzi, Taylor, Vu and Wong (2016) generated a multivariate Tweedie distribution for a claim triangle by means of common shocks. In a subsequent publication, the same authors (2021) noted that, if a common shock model were constructed by adding a multiple of a single common shock to each cell of a triangle, then it might be found to contribute proportionately heavily to some cells and only lightly to others. Such triangles were said to be **unbalanced** (with respect to the shocks). Equally, it might have been said that the common shock model was unbalanced with respect to the data.

Those authors defined, in response, a procedure that tended to equalize the common shock proportionate contributions over the cells of the triangle. The common shock model was then **balanced**, or at least more balanced.

The present paper continues to work within the algebra of common shocks in the general setting of Avanzi, Taylor and Wong (2018), and identifies certain models that are **auto-balanced**, i.e. within themselves, balanced without any adjustment by an equalization procedure.

Section 2 covers some preliminaries relating to the Tweedie family of distributions and common shock models in generality. Section 3 discusses the application of these general common shock models to Tweedie distributed shocks and idiosyncratic components, and Section 4 derives the first two moments of observations in these models.

Section 5 derives conditions that define the subset of these Tweedie common shock models that are auto-balanced. A number of numerical illustration of auto-balance models are set out in Section 6, and Section 7 then concludes.

## 2. Framework and notation

### 2.1. Notation

Although the papers of Avanzi, Taylor, Vu and Wong (2016, 2020) on balance worked in the context of claim triangles, the present paper will adopt the more general framework of Avanzi, Taylor and Wong (2018).

A claim array $\mathcal{A}$ will be defined here as a 2-dimensional array of random variables $X_{ij}$, indexed by integers $i, j$, with $1 \leq i \leq I, 1 \leq j \leq J$ for some fixed integers $I, J$. For any given pair $i, j$, the random variable $X_{ij}$ may or may not be present.

The subscripts $i, j$ typically index accident period (row) and development period (column) respectively, and the $X_{ij}$ represent observations on claims, commonly claim counts or amounts. In the special case $I = J$ and $\mathcal{A} = \{X_{ij}: 1 \leq i \leq I, 1 \leq j \leq I - i + 1\}$, the array reduces to the well known claim triangle.

Define $t = i + j - 1$, so that $t = 1, 2, \ldots, I + J - 1$. Observations with common $t$ lie on the $t$-th diagonal of $\mathcal{A}$.

Subsequent sections will often involve the simultaneous consideration of multiple business segments, with one array for each segment. A segment could be a line of business.

It will be necessary in this case to consider a collection $\mathbb{A} = \{\mathcal{A}^{(n)}, n = 1, 2, \ldots, N\}$ of claim arrays, where $\mathcal{A}^{(n)}$ denotes the array for segment $n$. It will be assumed initially that all $\mathcal{A}^{(n)}$ are congruent, i.e. are of the same dimensions $I, J$, and that they have missing observations in the same $i, j$ locations, but this assumption will be relaxed in Section 5.2.

The $i, j$ observation of $\mathcal{A}^{(n)}$ will be denoted $X_{ij}^{(n)}$; the entire $i$-th row of $\mathcal{A}^{(n)}$ denoted $\mathcal{R}_i^{(n)}$; and the the entire $j$-th column $\mathcal{C}_j^{(n)}$.

It will also be useful to consider diagonals of $\mathcal{A}^{(n)}$, where the $t$-th diagonal is defined as the subset $\{X_{ij}^{(n)} \in \mathcal{A}^{(n)}: i + j - 1 = t\}$, and represents claim observations from the $t$-th calendar period, $t = 1$ denoting the calendar period in which the first accident period falls. The entire $t$-th diagonal of $\mathcal{A}^{(n)}$ will be denoted $\mathcal{D}_t^{(n)}$.

### 2.2. Tweedie family of distributions

#### 2.2.1. Univariate Tweedie

The Tweedie family is a sub-family of the **exponential dispersion family ("EDF")**. The latter has two well known representations, the additive and reproductive forms (Jorgensen, 1987). In common with Avanzi, Taylor, Vu and Wong (2016), the present note commences within the context of the additive representation, which has the pdf

$$p(X = x; \theta, \lambda) = a(x, \phi) exp\{x\theta - \lambda b(\theta)\}, \tag{2.1}$$

where $\theta$ is a canonical parameter, $\lambda > 0$ is a index parameter, and $b(\theta)$ is called the cumulant function.

This distribution has cumulant generating function

$$K_X(t) = \lambda[b(\theta + t) - b(\theta)], \tag{2.2}$$

giving

$$E[X] = \lambda b'(\theta), \tag{2.3}$$

$$Var[X] = \lambda b''(\theta). \tag{2.4}$$

The **Tweedie sub-family** is obtained by the selection

$$b(\theta) = b_p(\theta) = \frac{1}{2-p}[(1-p)\theta]^{\frac{2-p}{1-p}}, p\epsilon(-\infty, 0] \cup (1,\infty), p \neq 1,2,$$
$$= \exp\theta, p = 1,$$
$$= -\ln(-\theta), p = 2. \tag{2.5}$$

If $X$ is distributed according to (2.1) with $b(\theta) = b_p(\theta)$, then it will be denoted $X \sim Tw_p^*(\theta, \lambda)$.

A useful alternative form of (2.5) in the case $p \neq 1,2$ is

$$b_p(\theta) = \frac{\alpha - 1}{\alpha}\left[\frac{\theta}{\alpha - 1}\right]^\alpha, \tag{2.6}$$

where $\alpha = (2 - p)/(1 - p)$.

**Remark 2.1.** It is required that $sgn(\theta) = sgn(\alpha - 1)$ if (2.6) is to be defined for $\alpha$ negative or fractional. ∎

**Remark 2.2.** It follows from (2.3)-(2.6) that, for $Tw_p^*$ with $p \neq 1$,

$$E[X] = \lambda\left[\frac{\theta}{\alpha - 1}\right]^{\alpha - 1}. \tag{2.7}$$

$$Var[X] = \lambda\left[\frac{\theta}{\alpha - 1}\right]^{\alpha - 2}, \tag{2.8}$$

and so

$$\theta = (\alpha - 1)\frac{E[X]}{Var[X]}. \tag{2.9}$$

It now follows from (2.7) and (2.9) that

$$\frac{1}{CoV^2[X]} = E[X]\frac{E[X]}{Var[X]} = \lambda \left[\frac{\theta}{\alpha-1}\right]^\alpha, \tag{2.10}$$

where $CoV[X]$ denotes the coefficient of variation of $X$.

It may be checked by means of (2.5) that (2.7), (2.8) and (2.10) continue to hold when $p = 1$ provided that the right hand sides are interpreted as the limiting case as $\alpha \to \infty$. This yields.

$$E[X] = Var[X] = \frac{1}{CoV^2[X]} = \lambda e^\theta \text{ for } p = 1. \tag{2.11}$$

Thus, for fixed $\alpha$, i.e. fixed $p$, Tweedie distributions with the same $\theta$ are those with the same mean-to-variance ratio. ∎

It will sometimes be found useful in subsequent development to re-parameterize $Tw_p^*(\theta, \lambda)$ in terms of $(\mu, v)$, where $\mu = E[X], v = CoV^2[X]$. It will also be useful to denote $\sigma^2 = Var[X] = \mu^2 v$. The re-parameterization is as set out in the following lemma. The alternative parameterizations are essentially related to the additive and reproductive representation of the EDF, and these representations of the Tweedie sub-family are also found in Avanzi, Taylor, Vu and Wong (2016).

**Lemma 2.3.** If $X \sim Tw_p^*(\theta, \lambda)$, then it may be re-parameterized as $X \sim Tw_p^*\left(\frac{\alpha-1}{\mu v}, \mu^\alpha v^{\alpha-1}\right)$.

**Proof.** See Appendix A. ∎

The following lemma derives some closure properties of the Tweedie family under scaling and addition of variates.

**Lemma 2.4.** Suppose that, for stochastically independent variates $V_1, V_2, V_i \sim Tw_p^*(\theta, \lambda_i), i = 1,2$. Then

$$kV_i \sim Tw_p^*(\theta/k, \lambda_i k^\alpha), i = 1,2 \text{ for constant } k > 0 \text{ and } p \neq 1, \tag{2.12}$$

$$V_1 + V_2 \sim Tw_p^*(\theta, \lambda_1 + \lambda_2). \tag{2.13}$$

**Proof.** By simple manipulation of the cgf (2.2), using (2.5) and (2.6). The results can also be found in Jørgensen (1997). ∎

2.2.2. Multivariate Tweedie
Following Furman and Landsman (2010), Avanzi, Taylor, Vu and Wong (2016) consider variates of the form

$$X_{ij}^{(n)} = \frac{\theta}{\theta_{ij}^{(n)}} W_{ij} + Z_{ij}^{(n)}, \tag{2.14}$$

where $W_{ij} \sim Tw_p^*(\theta, \lambda), Z_{ij}^{(n)} \sim Tw_p^*\left(\theta_{ij}^{(n)}, \lambda_{ij}^{(n)}\right)$ with $\theta_{ij}^{(n)} = \theta$ when $p = 1$.

Then

$$X_{ij}^{(n)} \sim Tw_p^*\left(\theta, \lambda + \lambda_{ij}^{(n)}\right) \text{ for } p = 1, \quad (2.15)$$

$$X_{ij}^{(n)} \sim Tw_p^*\left(\theta_{ij}^{(n)}, \lambda \left(\frac{\theta}{\theta_{ij}^{(n)}}\right)^\alpha + \lambda_{ij}^{(n)}\right) \text{ for } p \neq 1. \quad (2.16)$$

These results may be checked against Lemma 2.4. Since the marginals are Tweedie, the multi-dimensional variate $\left(X_{ij}^{(1)}, \ldots, X_{ij}^{(N)}\right)$ is **multivariate Tweedie**. It should be noted that, if the multiplier of $W_{ij}$ is non-zero and other than that shown in (2.14), then $X_{ij}^{(n)}$ is not Tweedie unless $p = 0$ (normal distribution).

It is interesting to take the $(\mu, v)$-parameterization of (2.16), using Lemma 2.3, to obtain

$$X_{ij}^{(n)} \sim Tw_p^*\left(\frac{\alpha - 1}{\mu_{ij}^{(n)} v_{ij}^{(n)}}, \left(\mu_{ij}^{(n)} v_{ij}^{(n)}\right)^\alpha \left(\frac{1}{v} + \frac{1}{v_{ij}^{(n)}}\right)\right) \text{ for } p \neq 1, \quad (2.17)$$

where the suffixes on $\mu, v$ correspond to those on $\theta, \lambda$.

## 2.3. Common shock models

The general common shock framework of Avanzi, Taylor and Wong (2018) is as follows.

Let $\mathcal{P}^{(n)}$ be a partition of $\mathcal{A}^{(n)} \in \mathbb{A}$, i.e. $\mathcal{P}^{(n)} = \left\{\mathcal{P}_1^{(n)}, \ldots, \mathcal{P}_P^{(n)}\right\}$ where the $\mathcal{P}_p^{(n)}$ are subsets of $\mathcal{A}^{(n)}$ with $\mathcal{P}_p^{(n)} \cap \mathcal{P}_q^{(n)} = \emptyset$ for all $p, q = 1, \ldots, P, p \neq q$ and $\bigcup_{p=1}^{P} \mathcal{P}_p^{(n)} = \mathcal{A}^{(n)}$. Suppose that all partitions are the same in the sense that, for each $p$, the $(i, j)$ positions of the elements of $\mathcal{A}^{(n)}$ included in $\mathcal{P}_p^{(n)}$ are the same for different $n$.

Now consider the following dependency structure on the elements $X_{ij}^{(n)}$:

$$X_{ij}^{(n)} = \alpha_{ij}^{(n)} W_{\pi(i,j)} + \beta_{ij}^{(n)} W_{\pi(i,j)}^{(n)} + \phi_{ij}^{(n)} Z_{ij}^{(n)} \quad (2.18)$$

where $\pi(i, j) = p$ such that $X_{ij}^{(n)} \in \mathcal{P}_p^{(n)}$, a unique mapping; $W_{\pi(i,j)}, W_{\pi(i,j)}^{(n)}, Z_{ij}^{(n)}$ are independent stochastic variates, and $\alpha_{ij}^{(n)}, \beta_{ij}^{(n)}, \phi_{ij}^{(n)} \geq 0$ are fixed and known constants (**"mixture constants"**); a convenient terminology for the $W_{\pi(i,j)}, W_{\pi(i,j)}^{(n)}, Z_{ij}^{(n)}$ comprises **umbrella common shocks, array-specific common shocks**, and **idiosyncratic components** respectively. The $\pi(i, j)$ will be referred to as **partition subsets**.

It is evident that $W_p$ is a common shock across all $n$, but affecting only subsets $\mathcal{P}_p^{(n)}$ for fixed $p$; $W_p^{(n)}$ is similarly a common shock across $\mathcal{P}_p^{(n)}$, but now for fixed $n$ and $p$; and $Z_{ij}^{(n)}$ is an idiosyncratic component of $X_{ij}^{(n)}$, specific to $i, j$ and $n$.

The common shock $W_p^{(n)}$ creates dependency between observations within the subset $\mathcal{P}_p^{(n)}$ of array $\mathcal{A}^{(n)}$. Since the partitions $\mathcal{P}^{(n)}$ are the same across $n$, the common shock $W_p$ creates dependency between observations in the subsets $\mathcal{P}_p^{(n)}$ of the same or different arrays.

Particular selections of partitions $\mathcal{P}^{(n)}$ are of special interest, as set out in Table 2-1.

**Table 2-1 Special cases of common shock**

| Type of dependence | Partition $\mathcal{P}^{(n)}$ |
|---|---|
| Array-wide | $\{\mathcal{P}_1^{(n)}\}$ with $\mathcal{P}_1^{(n)} = \mathcal{A}^{(n)}$ |
| Cell-wise | $\{\mathcal{P}_1^{(n)}, \mathcal{P}_2^{(n)}, \ldots\}$ with $\mathcal{P}_1^{(n)} = \{X_{11}^{(n)}\}, \mathcal{P}_2^{(n)} = \{X_{12}^{(n)}\}, \ldots$ |
| Row-wise | $\{\mathcal{P}_1^{(n)}, \ldots, \mathcal{P}_I^{(n)}\}$ with $\mathcal{P}_i^{(n)} = \mathcal{R}_i^{(n)}$ |
| Column-wise | $\{\mathcal{P}_1^{(n)}, \ldots, \mathcal{P}_J^{(n)}\}$ with $\mathcal{P}_j^{(n)} = \mathcal{C}_j^{(n)}$ |
| Diagonal-wise | $\{\mathcal{P}_1^{(n)}, \ldots, \mathcal{P}_{I+J-1}^{(n)}\}$ with $\mathcal{P}_t^{(n)} = \mathcal{D}_t^{(n)}$ |

**Remark 2.5.** Since multiplication of a $Tw_p^*$ variate by a constant produces another $Tw_p^*$ variate for $p \neq 1$ (Lemma 2.4), the multiplier $\phi_{ij}^{(n)}$ in (2.18) may be absorbed into $Z_{ij}^{(n)}$ so that (2.18) simplifies to

$$X_{ij}^{(n)} = \alpha_{ij}^{(n)} W_{\pi(i,j)} + \beta_{ij}^{(n)} W_{\pi(i,j)}^{(n)} + Z_{ij}^{(n)} \text{ for } p \neq 1. \tag{2.19} \blacksquare$$

This will be taken as the general common shock structure for the remainder of this paper.

## 3. Multivariate Tweedie for a general common shock model

The cell structure (2.14) used by Avanzi, Taylor, Vu and Wong (2016) is a special case of the general common shock formula (2.19) with $\mathcal{P}^{(n)}$ as in the cell-wise example of Table 2-1, $\alpha_{ij}^{(n)} = \theta/\theta_{ij}^{(n)}, \beta_{ij}^{(n)} = 0, \phi_{ij}^{(n)} = 1$. Since (2.14) generates a multivariate Tweedie, the objective will now be to identify the most general case of (2.19) that also generates a multivariate Tweedie. This is done in Appendix A, leading to the following result.

**Lemma 3.1.** Suppose that $W_{\pi(i,j)} \sim Tw_p^*(\theta_{\pi(i,j)}, \lambda_{\pi(i,j)})$, $W_{\pi(i,j)}^{(n)} \sim Tw_p^*(\theta_{\pi(i,j)}^{(n)}, \lambda_{\pi(i,j)}^{(n)})$, $Z_{ij}^{(n)} \sim Tw_p^*(\theta_{ij}^{(n)}, \lambda_{ij}^{(n)})$ with $\theta_{\pi(i,j)} = \theta_{\pi(i,j)}^{(n)} = \theta_{ij}^{(n)}$ in the case $p = 1$. Then $X_{ij}^{(n)}$ is Tweedie distributed if and only if $\alpha_{ij}^{(n)} = \theta_{\pi(i,j)}/\theta_{ij}^{(n)}, \beta_{ij}^{(n)} = \theta_{\pi(i,j)}^{(n)}/\theta_{ij}^{(n)}$. In fact,

$$X_{ij}^{(n)} \sim Tw_p^*\left(\theta_{ij}^{(n)}, \lambda_{\pi(i,j)} + \lambda_{\pi(i,j)}^{(n)} + \lambda_{ij}^{(n)}\right) \text{ for } p = 1. \tag{3.1}$$

$$X_{ij}^{(n)} \sim Tw_p^*\left(\theta_{ij}^{(n)}, \left(\frac{\theta_{\pi(i,j)}}{\theta_{ij}^{(n)}}\right)^\alpha \lambda_{\pi(i,j)} + \left(\frac{\theta_{\pi(i,j)}^{(n)}}{\theta_{ij}^{(n)}}\right)^\alpha \lambda_{\pi(i,j)}^{(n)} + \lambda_{ij}^{(n)}\right) \text{ for } p \neq 1. \tag{3.2} \blacksquare$$

**Corollary 3.2.** In the case $p = 2$ (i.e. $\alpha = 0$), the result (3.2) reduces to (3.1). $\blacksquare$

The following result is obtained by re-parameterization of Lemma 3.1 according to Lemma 2.3.

**Corollary 3.3.** In Lemma 3.1, the coefficients $\alpha_{ij}^{(n)}, \beta_{ij}^{(n)}$ may be re-parameterized as

$$\alpha_{ij}^{(n)} = \frac{\mu_{ij}^{(n)}}{\mu_{\pi(i,j)}} \frac{v_{ij}^{(n)}}{v_{\pi(i,j)}}, \tag{3.3}$$

$$\beta_{ij}^{(n)} = \frac{\mu_{ij}^{(n)}}{\mu_{\pi(i,j)}^{(n)}} \frac{v_{ij}^{(n)}}{v_{\pi(i,j)}^{(n)}}, \tag{3.4}$$

and the result (3.2) may be re-parameterized as

$$X_{ij}^{(n)} \sim Tw_p^* \left( \frac{\alpha-1}{\mu_{ij}^{(n)} v_{ij}^{(n)}}, \left( \mu_{ij}^{(n)} v_{ij}^{(n)} \right)^\alpha \left( \frac{1}{v_{\pi(i,j)}} + \frac{1}{v_{\pi(i,j)}^{(n)}} + \frac{1}{v_{ij}^{(n)}} \right) \right) \text{ for } p \ne 1, \tag{3.5} \blacksquare$$

**Corollary 3.4.** The mixture coefficients $\alpha_{ij}^{(n)}, \beta_{ij}^{(n)}$ do not depend on $p$ (or $\alpha$). $\blacksquare$

The model obtained from the general common shock model (2.19) by the substitutions of Lemma 3.1 or Corollary 3.3 for $\alpha_{ij}^{(n)}, \beta_{ij}^{(n)}$ will be referred to as the **general multivariate Tweedie common shock model**.

Henceforth, results will be expressed in the $(\mu, v)$ parameterization.

**Example 3.5.** Corollary 3.3 may be illustrated for the case $p \ne 1$ with the choice of row-wise dependence from Table 2-1, where $\mathcal{P}_i^{(n)} = \mathcal{R}_i^{(n)}$, and so $\pi(i,j) = i$. In this case, (2.19) and (3.5) become

$$X_{ij}^{(n)} = \left( \frac{\mu_{ij}^{(n)}}{\mu_i} \frac{v_{ij}^{(n)}}{v_i} \right) W_i + \left( \frac{\mu_{ij}^{(n)}}{\mu_i^{(n)}} \frac{v_{ij}^{(n)}}{v_i^{(n)}} \right) W_i^{(n)} + Z_{ij}^{(n)} \tag{3.6}$$

and

$$X_{ij}^{(n)} \sim Tw_p^* \left( \frac{\alpha-1}{\mu_{ij}^{(n)} v_{ij}^{(n)}}, \left( \mu_{ij}^{(n)} v_{ij}^{(n)} \right)^\alpha \left( \frac{1}{v_i} + \frac{1}{v_i^{(n)}} + \frac{1}{v_{ij}^{(n)}} \right) \right), \tag{3.7}$$

with $W_i \sim Tw_p^* \left( \frac{\alpha-1}{\mu_i v_i}, \mu_i^\alpha v_i^{\alpha-1} \right)$, $W_i^{(n)} \sim Tw_p^* \left( \frac{\alpha-1}{\mu_i^{(n)} v_i^{(n)}}, \left( \mu_i^{(n)} \right)^\alpha \left( v_i^{(n)} \right)^{\alpha-1} \right)$.

According to (3.6), the observation in the $(i, j)$ cell consists of three contributions:

(1) a shock that impacts all cells of row $i$ in all arrays;
(2) a shock that impacts all cells of row $i$ in just the $n$-th array;
(3) an idiosyncratic component that impacts just the $(i, j)$ cell.

All three components are $Tw_p^*$ distributed, as is the total observation, and all have the same canonical parameter $\theta_{ij}^{(n)} = \frac{\alpha-1}{\mu_{ij}^{(n)} v_{ij}^{(n)}}$. $\blacksquare$

# 4. Moments of observations in a multivariate Tweedie common shock model

Consider the general multivariate Tweedie common shock model, with observations represented by (3.1) or (3.2). The first two moments of this observation are as in the following lemma.

**Lemma 4.1.** For $X_{ij}^{(n)}$ an observation in a general multivariate Tweedie common shock model,

$$E\left[X_{ij}^{(n)}\right] = \mu_{ij}^{(n)} \left( \frac{v_{ij}^{(n)}}{v_{\pi(i,j)}} + \frac{v_{ij}^{(n)}}{v_{\pi(i,j)}^{(n)}} + 1 \right), \tag{4.1}$$

$$Var\left[X_{ij}^{(n)}\right] = \left(\sigma_{ij}^{(n)}\right)^2 \left( \frac{v_{ij}^{(n)}}{v_{\pi(i,j)}} + \frac{v_{ij}^{(n)}}{v_{\pi(i,j)}^{(n)}} + 1 \right). \tag{4.2}$$

Note that these results hold without restriction on $p$. ∎

**Proof.** See Appendix A.

The result just below then follows.

**Proposition 4.2.** For a general multivariate Tweedie common shock model,

(a) $E\left[X_{ij}^{(n)}\right]$ and $Var\left[X_{ij}^{(n)}\right]$ are the same multiples of $E\left[Z_{ij}^{(n)}\right]$ and $Var\left[Z_{ij}^{(n)}\right]$ respectively;
(b) The common multiple decomposes into its two common shock and single idiosyncratic components in the same proportions for $E\left[X_{ij}^{(n)}\right]$ and $Var\left[X_{ij}^{(n)}\right]$ respectively. The multiples do not depend on $p$ (or $\alpha$). ∎

# 5. Balance of a general multivariate Tweedie common shock model

## 5.1. Condition for auto-balance

As foreshadowed in Section 1, a common shock model is regarded as balanced if the proportionate contribution of each shock and idiosyncratic component to cell expectation is constant over all cells. Now, in the case of general multivariate Tweedie common shock model, the cell expectation is given by (4.1), where the three components within the bracket correspond to the two common shock and the idiosyncratic contributions respectively.

Hence auto-balance occurs if and only if, for each $n$, the ratios $v_{ij}^{(n)}/v_{\pi(i,j)}$ and $v_{ij}^{(n)}/v_{\pi(i,j)}^{(n)}$ are independent of $i, j$. This leads to the necessary and sufficient condition set out in the following proposition.

**Proposition 5.1.** A general multivariate Tweedie common shock model will be auto-balanced if and only if the following two conditions hold:

(a) $v_{ij}^{(n)} = C^{(n)} v_{\pi(i,j)}$; and

(b) $v_{ij}^{(n)} = K^{(n)} v_{\pi(i,j)}^{(n)}$,

for all $i, j, n$, where $C^{(n)}, K^{(n)} > 0$ are quantities that depend on only $n$.

For this auto-balanced case,

$$E\left[X_{ij}^{(n)}\right] = \kappa^{(n)} \mu_{ij}^{(n)}, \tag{5.1}$$

$$Var\left[X_{ij}^{(n)}\right] = \kappa^{(n)} \left(\sigma_{ij}^{(n)}\right)^2, \tag{5.2}$$

$$CoV^2\left[X_{ij}^{(n)}\right] = \left(\kappa^{(n)}\right)^{-1} v_{ij}^{(n)}, \tag{5.3}$$

where $\kappa^{(n)} = K^{(n)} + C^{(n)} + 1$, which does not depend on $p$ (or $\alpha$).

**Proof.** See Appendix A. ∎

The following remark illustrates that, despite the restrictions of Proposition 5.1 on CoVs, the auto-balanced general multivariate Tweedie common shock model retains richness of structure.

**Remark 5.2.** The CoVs of umbrella common shocks for partition subsets are unrestricted. For a particular array, the CoVs of the array-specific common shocks over partition subsets must be common multiples of the umbrella common shocks for the corresponding partition subsets. These CoVs may thus vary by both array and partition subset. Within each array, the idiosyncratic CoVs must be common multiples of the CoVs of the array-specific common shocks for the corresponding partition subsets. Thus, the idiosyncratic CoVs may also vary by both array and partition subset, but not within a partition subset of an array. Under these constraints, the total cell CoVs within a particular array are common multiples of the idiosyncratic CoVs. The multiples may vary by array. ∎

**Remark 5.3.** For the model of Proposition 5.1, the expected cell values $E\left[X_{ij}^{(n)}\right]$ are all constant multiples of the idiosyncratic expectations $\mu_{ij}^{(n)}$, where the multiples depend on only the array number $n$. Likewise, cell variances $Var\left[X_{ij}^{(n)}\right]$ are the same constant multiples of the squared idiosyncratic variances $\left[\sigma_{ij}^{(n)}\right]^2$. ∎

There may be occasions on which one wishes to omit either the umbrella or array-specific common shocks from the model of Lemma 3.1. In such cases, the proof of Proposition 5.1 in Appendix A is simply modified to obtain the following result.

**Remark 5.4.** Suppose that the umbrella common shock is omitted from the general multivariate Tweedie common shock model for particular array $n = n^*$. Then Proposition 5.1 holds with condition (a) omitted for $n = n^*$, and $C^{(n^*)}$ omitted from $\kappa^{(n^*)}$. Similarly, if the array-specific common shock is omitted for array $n = n^*$, then the proposition holds with condition (b) omitted for $n = n^*$, and $K^{(n^*)}$ omitted from $\kappa^{(n^*)}$. ∎

## 5.2. Model extensions

The general multivariate Tweedie common shock model set out in Lemma 3.1 contains two common shock components, the umbrella and array-specific, and both affect the same partition subsets $\pi(i,j)$. The model can be generalized by adding further shocks and allowing different shocks to affect different subsets.

Section 2.3 introduced the partition $\mathcal{P}^{(n)}$ of the claim array $\mathcal{A}^{(n)}$. Of course, the array may be partitioned in various ways, e.g. accident years, calendar years, etc., as illustrated in Table 2-1, and a set of common shocks associated with each partition.

Accordingly, consider $R$ distinct partitions $\mathcal{P}^{(n)}_{[r]}, r = 1, \ldots, R$ of each array $\mathcal{A}^{(n)}$. Subsets of $\mathcal{P}^{(n)}_{[r]}$ will be denoted by $\mathcal{P}^{(n)}_{[r]1}, \mathcal{P}^{(n)}_{[r]2}, \ldots$. For each $r$, let $\pi_{[r]}(i,j)$ denote the partition subset that contains cell $(i,j)$. Next generalize (2.19) to the following:

$$X^{(n)}_{ij} = \sum_{r=1}^{R} \alpha^{(n)}_{ij,[r]} W_{\pi_{[r]}(i,j)} + \sum_{r=1}^{R} \beta^{(n)}_{ij,[r]} W^{(n)}_{\pi_{[r]}(i,j)} + Z^{(n)}_{ij}, \tag{5.4}$$

where $W_{\pi_{[r]}(i,j)}, W^{(n)}_{\pi_{[r]}(i,j)}$ are umbrella and array-specific common shocks affecting partition subset $\pi_{[r]}(i,j)$ and all $W_{\pi_{[r]}(i,j)}, W^{(n)}_{\pi_{[r]}(i,j)}, Z^{(n)}_{ij}$ are independent. Adopt the notation $v_{\pi_{[r]}(i,j)} = CoV^2\left[W_{\pi_{[r]}(i,j)}\right], v^{(n)}_{\pi_{[r]}(i,j)} = CoV^2\left[W^{(n)}_{\pi_{[r]}(i,j)}\right]$.

By convention, and despite the $\Sigma_{r=1}^R$ notation, it will be permitted that some of the variates $W_{\pi_{[r]}(i,j)}, W^{(n)}_{\pi_{[r]}(i,j)}$ may be absent from (5.4), but it is assumed that at least one or the other is present for each $r$. It will be assumed that the same terms $W_{\pi_{[r]}(i,j)}, W^{(n)}_{\pi_{[r]}(i,j)}$ are present for each $i,j,n$. Let $\chi$ and $\chi^{(n)}$ respectively denote the number of $W_{\pi_{[r]}(i,j)}$ and $W^{(n)}_{\pi_{[r]}(i,j)}$ terms present.

**Example 5.5.** A possible model is as follows:

$$X^{(n)}_{ij} = \alpha^{(n)}_{ij,[1]} W_{\pi_{[1]}(i,j)} + \beta^{(n)}_{ij,[1]} W^{(n)}_{\pi_{[1]}(i,j)} + \beta^{(n)}_{ij,[2]} W^{(n)}_{\pi_{[2]}(i,j)} + Z^{(n)}_{ij}, \tag{5.5}$$

where the $\pi_{[r]}(i,j)$ are defined by $\mathcal{P}^{(n)}_{[1]}$, the diagonal-wise partition and $\mathcal{P}^{(n)}_{[2]}$, the row-wise partition from Table 2-1. The model thus includes a diagonal-wise umbrella common shock, and both diagonal-wise and row-wise array-specific common shocks. ∎

Lemma 3.1 and Corollary 3.3 extend easily to this more general situation, as set out in the following proposition.

**Proposition 5.6.** Suppose that, in the model (5.4), $W_{\pi_{[r]}(i,j)} \sim Tw_p^*\left(\theta_{\pi_{[r]}(i,j)}, \lambda_{\pi_{[r]}(i,j)}\right)$, $W^{(n)}_{\pi_{[r]}(i,j)} \sim Tw_p^*\left(\theta^{(n)}_{\pi_{[r]}(i,j)}, \lambda^{(n)}_{\pi_{[r]}(i,j)}\right)$, $Z^{(n)}_{ij} \sim Tw_p^*\left(\theta^{(n)}_{ij}, \lambda^{(n)}_{ij}\right)$ with $\theta_{\pi_{[r]}(i,j)} = \theta^{(n)}_{\pi_{[r]}(i,j)} = \theta^{(n)}_{ij}$ in the case $p = 1$. Then $X^{(n)}_{ij}$ is Tweedie distributed if and only if $\alpha^{(n)}_{ij,[r]} = \theta_{\pi_{[r]}(i,j)}/\theta^{(n)}_{ij}, \beta^{(n)}_{ij,[r]} = \theta^{(n)}_{\pi_{[r]}(i,j)}/\theta^{(n)}_{ij}$. In fact,

$$X_{ij}^{(n)} \sim Tw_p^*\left(\theta_{ij}^{(n)}, \sum_{r=1}^{R} \lambda_{\pi_{[r]}(i,j)} + \sum_{r=1}^{R} \lambda_{\pi_{[r]}(i,j)}^{(n)} + \lambda_{ij}^{(n)}\right) \text{ for } p = 1. \tag{5.6}$$

$$X_{ij}^{(n)} \sim Tw_p^*\left(\theta_{ij}^{(n)}, \sum_{r=1}^{R} \left(\frac{\theta_{\pi_{[r]}(i,j)}^{(n)}}{\theta_{ij}^{(n)}}\right)^{\alpha} \lambda_{\pi_{[r]}(i,j)} + \sum_{r=1}^{R} \left(\frac{\theta_{\pi_{[r]}(i,j)}^{(n)}}{\theta_{ij}^{(n)}}\right)^{\alpha} \lambda_{\pi_{[r]}(i,j)}^{(n)}\right.$$
$$\left. + \lambda_{ij}^{(n)}\right) \text{ for } p \neq 1. \tag{5.7}$$

As in Corollary 3.3, these results may be expressed in the alternative form:

$$\alpha_{ij,[r]}^{(n)} = \frac{\mu_{ij}^{(n)}}{\mu_{\pi_{[r]}(i,j)}^{(n)}} \frac{v_{ij}^{(n)}}{v_{\pi_{[r]}(i,j)}^{(n)}}, \tag{5.8}$$

$$\beta_{ij,[r]}^{(n)} = \frac{\mu_{ij}^{(n)}}{\mu_{\pi_{[r]}(i,j)}^{(n)}} \frac{v_{ij}^{(n)}}{v_{\pi_{[r]}(i,j)}^{(n)}}, \tag{5.9}$$

$$X_{ij}^{(n)} \sim Tw_p^*\left(\frac{\alpha-1}{\mu_{ij}^{(n)} v_{ij}^{(n)}}, \left(\mu_{ij}^{(n)} v_{ij}^{(n)}\right)^{\alpha} \left(\sum_{r=1}^{R} \frac{1}{v_{\pi_{[r]}(i,j)}} + \sum_{r=1}^{R} \frac{1}{v_{\pi_{[r]}(i,j)}^{(n)}} + \frac{1}{v_{ij}^{(n)}}\right)\right) \text{ for } p$$
$$\neq 1, \tag{5.10} \blacksquare$$

The establishment of conditions for auto-balance requires the notion of connectedness of cells of an array. Cells $(i,j)$ and $(k,\ell)$ of array $\mathcal{A}^{(n)}$ will be said to be **connected** if there exists a sequence of subsets $\left\{\mathcal{P}_{[r_g]s_g}^{(n)}, g = 1, \dots, G\right\}$ such that $(i,j) \in \mathcal{P}_{[r_1]s_1}^{(n)}, (k,\ell) \in \mathcal{P}_{[r_G]s_G}^{(n)}$ and $\mathcal{P}_{[r_g]s_g}^{(n)} \cap \mathcal{P}_{[r_{g+1}]s_{g+1}}^{(n)} \neq \emptyset, g = 1, \dots, G-1$. By congruence of the arrays $\mathcal{A}^{(n)}$, if $(i,j)$ and $(k,\ell)$ are connected within one array, they will be connected in all others.

It is evident that connectedness between two cells is an equivalence relation, and so each array $\mathcal{A}^{(n)}$ partitions into equivalence classes $\mathcal{E}_h, h = 1, \dots, H$ which, by congruence, do not depend on $n$. The cells within any one equivalence class will be connected to all other cells in the class, and disconnected from all cells in all other classes.

Further, since all cells within a single partition subset are connected to all other cells in that subset, an equivalence class must consist of a union of partition subsets. It will sometimes be convenient to denote an equivalence class $\mathcal{E}_h$ by $\mathcal{E}(i,j)$ for any cell $(i,j)$ contained in the class.

Proposition 5.7 now gives the necessary and sufficient condition for auto-balance in the case of model (5.4).

**Proposition 5.7.** Consider a Tweedie common shock model (5.4), subject to the parameter restrictions imposed by Proposition 5.6. The model will be auto-balanced if and only if the following two conditions hold:

(a) $v_{ij}^{(n)} = C^{(n)} v_{\pi_{[r]}(i,j)}$; and
(b) $v_{ij}^{(n)} = K^{(n)} v_{\pi_{[r]}(i,j)}^{(n)}$,

for all $i, j, n, r$, where $C^{(n)}, K^{(n)} > 0$ are quantities that depend on only $n$.

For this auto-balanced case,

$$E\left[X_{ij}^{(n)}\right] = \kappa^{(n)} \mu_{ij}^{(n)}, \tag{5.11}$$

$$Var\left[X_{ij}^{(n)}\right] = \kappa^{(n)} \left(\sigma_{ij}^{(n)}\right)^2, \tag{5.12}$$

$$CoV^2\left[X_{ij}^{(n)}\right] = \left(\kappa^{(n)}\right)^{-1} v_{ij}^{(n)}, \tag{5.13}$$

where $\kappa^{(n)} = K^{(n)} \chi + C^{(n)} \chi^{(n)} + 1$, which does not depend on $p$ (or $\alpha$).

**Proof.** A slight modification of the proof of Proposition 5.1. ∎

This result sets the relationships of the idiosyncratic components with the common shock components in the same cells. However, auto-balance of the Tweedie common shock model of Proposition 5.7 imposes constraints on the common shocks, as stated in the following result.

**Proposition 5.8.** The auto-balance conditions of Proposition 5.7 imply that, for any given $n, r, i, j$, $v_{\pi_{[r]}(k,\ell)}$ and $v_{\pi_{[r]}(k,\ell)}^{(n)}$ are constant (though not necessarily with each other) over all $(k, \ell) \in \mathcal{E}(i, j)$. In short, common shock CoVs are constant over equivalence classes.

**Proof.** See Appendix A. ∎

A few examples of connectedness in common shock models follow. As a preliminary, one may note that, in the case $R = 1$ in (5.4), there is a single partition of the arrays. The subsets of a partition are disjoint, by definition, and so there is no connectedness of cells in distinct subsets. The equivalence classes $\mathcal{E}_h$ are the partition subsets themselves. By Proposition 5.8, common shock CoVs are constant over just partition subsets, which is the case by their definition.

**Example 5.9: cell-wise common shocks.** The array partition for cell-wise shocks is given in Table 2-1, where it is seen that the partition subsets are the cells themselves. Hence, Proposition 5.8 states that common shock CoVs are constant over cells within a partition. This is a vacuous statement, which imposes no restriction on these CoVs. ∎

**Example 5.10: row-wise common shocks.** By the same type of argument as in Example 5.9, the common shock CoVs are constant over the partition subsets, which are rows. ∎

**Example 5.11: simultaneous row-wise and diagonal-wise common shocks.** This is the case $R = 2$ in (5.4), where the partition subsets $\pi_{[1]}(i, j)$ are rows and $\pi_{[2]}(k, \ell)$ diagonals.

It is possible to show that all $(i, j), (k, \ell)$ are connected. If $i = k$, then the two points are connected by virtue of lying within the same row. If $i + j = k + \ell$, then they are connected within the same diagonal. If neither of these conditions holds, then, without loss of generality, one may assume that $i + j < k + \ell$. One may connect the cell $(i, j)$ to $(k + \ell - i)$ in the same row, and thence to $(k, \ell)$ in the same diagonal.

Thus, the array contains only one equivalence class, namely the entire array. It then follows from Proposition 5.8 that the umbrella and array-specific common shock CoVs must be constant across all cells in each array if the model is to be balanced. ∎

**Example 5.12: a more exotic case.** In this case $R = 3$. Let $\mathcal{P}_{[1]}^{(n)}$ be the row-wise partition of Example 5.10, i.e. $\mathcal{P}_{[1]i}^{(n)} = \mathcal{R}_i^{(n)}$ (see Table 2-1). Now let $\mathcal{P}_{[*]}^{(n)}$ denote the diagonal-wise partition $\mathcal{P}_{[*]t}^{(n)} = \mathcal{D}_t^{(n)}$, and define $\mathcal{P}_{[2]}^{(n)}, \mathcal{P}_{[3]}^{(n)}$ as $\left\{\mathcal{D}_t^{(n)} \cap \mathcal{R}_i^{(n)} : i \leq i_0, all\ t\right\}$ and $\left\{\mathcal{D}_t^{(n)} \cap \mathcal{R}_i^{(n)} : i > i_0, all\ t\right\}$ respectively for some chosen $i_0$. Then $\mathcal{P}_{[2]}^{(n)}$ represents a diagonal effect in "low" accident periods, and $\mathcal{P}_{[3]}^{(n)}$ a diagonal effect in "high" accident periods.

In this case, rows are connected (as before), and the semi-diagonals constituting $\mathcal{P}_{[2]}^{(n)}$ are also connected, just as in Example 5.11. Similarly, the semi-diagonals in $\mathcal{P}_{[3]}^{(n)}$ are also connected. However, there is no connection between $\mathcal{P}_{[2]}^{(n)}, \mathcal{P}_{[3]}^{(n)}$, with the result that the equivalence classes are $\mathcal{P}_{[2]}^{(n)}, \mathcal{P}_{[3]}^{(n)}$. ∎

A further model extension is possible. Section 2.1 assumed congruence of all arrays $\mathcal{A}^{(n)}$, and this property has been useful in one or two of the proofs of auto-balance. Now that the conditions for auto-balance have been established, however, the assumption of congruence can be seen to be unnecessary.

**Proposition 5.13.** Consider the model of Proposition 5.6, which is sufficiently general to include the earlier model of Section 5.1. The conditions for auto-balance are given in Proposition 5.7, on the assumption of congruence of all arrays $\mathcal{A}^{(n)}$. Now delete arbitrary cells from any or all of the $\mathcal{A}^{(n)}$, not necessarily the same cells for each $n$. Then the model will be auto-balanced if and only conditions (a) and (b) of Proposition 5.7 hold for all $i, j, n, r$ for which the $v$ terms exist on both sides of the equations.

**Proof.** Consider the situation before the deletion of cells, and suppose that conditions (a) and (b) of Proposition 5.7 hold for all $i, j, n, r$, ensuring auto-balance. This means that the proportionate contribution of each shock and idiosyncratic component to cell expectation is constant over all cells observed. Now delete the nominated cells from each array. The auto-balance condition is unaffected in respect of the remaining cells. ∎

# 6. Numerical examples
## 6.1. Data sets

Three synthetic data sets were used to illustrate the auto-balance of three different claim models when conditions in Proposition 5.1 hold.

All data sets are Tweedie distributed with power parameter $p = 1.8$ and each data set consists of two triangles with dimension 15x15. Patterns over different development periods of the idiosyncratic and common shock components are different between triangles, and from each other. Parameters have been intentionally selected to reflect different degrees of contribution of umbrella and array-specific shocks within each triangle, and between triangles.

Data set 1 is used to illustrate the claim model with cell-wise common shocks described in Example 5.9. There is no restriction on the common shock CoVs as the partition subsets are cells themselves. Column-wise CoVs have been chosen only for the purpose of simplification.

The claim model with row-wise umbrella and array-specific shocks in Example 5.10 is illustrated using data set 2. The partition subsets are rows, and hence the model is specified with row-wise common shock CoVs.

The exotic case described in Example 5.12 is illustrated using the third data set. This data set has split diagonal-wise umbrella shocks which apply to two segments of accident periods $\{i \leq 10\}$ and $\{i > 10\}$. The array-specific shocks are row-wise. As described in Example 5.12, all cells within a row in an array are connected due to the array-specific shocks, and all cells within the same semi-diagonal are connected due to the split umbrella diagonal shocks. The two equivalence classes in each triangle are all cells with $\{i \leq 10\}$ and $\{i > 10\}$, respectively. CoVs of common shocks are specified to be constant for all cells within each of these equivalence classes.

Parameter specifications of these synthetic data sets are provided in Appendix B.

## 6.2. Common shock contributions

It is worthy of note that the conditions for auto-balance of common shock models aim to provide equality of expected proportionate contribution of each shock and idiosyncratic component to cell expectation over all cells. Hence, it is sufficient to demonstrate the auto-balance of common shock models using the expected value and CoV of each shock and idiosyncratic component. If one wishes to assess the balance of common shock contributions using simulated data, simulations could be carried out using the specified expected values, CoVs and power parameter $p$. This section provides the results of common shock contributions on both expected values and simulated values.

Following (2.19), (3.3) and (3.4), the expected proportionate contribution of umbrella and array-specific shocks to cell expectation overall all cells can be calculated by

$$\mu_{ij}^{(n)} \frac{v_{ij}^{(n)}}{v_{\pi(i,j)}} \Big/ E\left[X_{ij}^{(n)}\right],$$

and

$$\mu_{ij}^{(n)} \frac{v_{ij}^{(n)}}{v_{\pi(i,j)}^{(n)}} \Big/ E\left[X_{ij}^{(n)}\right],$$

respectively, and where $E\left[X_{ij}^{(n)}\right]$ is calculated using (4.1).

The expected common shock proportions for all three synthetic data sets arising from the parameter specifications in Appendix B are provided in Table 6-1.

Table 6-1 Expected common shock proportions to cell expectations of synthetic data

| Shock | Data set 1 – Cell-wise dependence | Data set 2 – Row-wise dependence | Data set 3 – Split diagonal umbrella shock and row-wise array-specific shock |
|---|---|---|---|
| Umbrella | Triangle 1: 11.4%, all cells<br>Triangle 2: 0.8%, all cells | Triangle 1: 3.6%, all cells<br>Triangle 2: 0.8%, all cells | Triangle 1: 3.9%, all cells<br>Triangle 2: 1.9%, all cells |
| Array-specific | Triangle 1: 1.0%, all cells<br>Triangle 2: 3.9%, all cells | Triangle 1: 0.7%, all cells<br>Triangle 2: 3.9%, all cells | Triangle 1: 0.9%, all cells<br>Triangle 2: 5.8%, all cells |

It is observed that each shock contributes equally to cell expectation across all cells and auto-balance is achieved. It can also be easily verified that the remaining contribution to cell expectation which belongs to the idiosyncratic component satisfies (5.1) with the shock multiples $C^{(n)}$ and $K^{(n)}$ provided in Appendix B.

For simulated data, the proportionate contribution of umbrella and array-specific shocks to cell total is calculated using the mixture coefficients and simulated values as

$$\frac{\mu_{ij}^{(n)}}{\mu_{\pi(i,j)}} \frac{v_{ij}^{(n)}}{v_{\pi(i,j)}} W_{\pi(i,j)} \Big/ X_{ij}^{(n)},$$

and

$$\frac{\mu_{ij}^{(n)}}{\mu_{\pi(i,j)}^{(n)}} \frac{v_{ij}^{(n)}}{v_{\pi(i,j)}^{(n)}} W_{\pi(i,j)}^{(n)} \Big/ X_{ij}^{(n)},$$

respectively, and where $X_{ij}^{(n)}$ is the sum of simulated components.

Table 6-2 and Table 6-3 show the contributions of umbrella shock and array-specific shock to cell total for triangle 1 in data set 3. The equivalent tables for the remaining data sets and triangles are provided in Appendix B.

Table 6-2 Proportionate contributions of umbrella shock to cell total – Triangle 1, data set 3

| Accident period | Development period | | | | | | | | | | | | | | |
|---|---|---|---|---|---|---|---|---|---|---|---|---|---|---|---|
| | 1 | 2 | 3 | 4 | 5 | 6 | 7 | 8 | 9 | 10 | 11 | 12 | 13 | 14 | 15 |
| 1 | 0.7% | 4.9% | 3.7% | 2.1% | 8.3% | 4.9% | 0.5% | 3.1% | 0.8% | 6.9% | 4.5% | 21.7% | 2.5% | 3.3% | 0.6% |
| 2 | 5.1% | 3.2% | 2.5% | 11.2% | 3.8% | 0.5% | 3.0% | 0.9% | 3.0% | 5.3% | 21.1% | 1.7% | 2.0% | 1.0% | |
| 3 | 2.8% | 2.3% | 8.2% | 3.2% | 0.4% | 1.9% | 1.0% | 2.5% | 5.3% | 11.7% | 2.9% | 2.6% | 1.0% | | |
| 4 | 4.6% | 4.2% | 3.3% | 0.4% | 2.2% | 1.6% | 3.1% | 6.4% | 16.0% | 2.0% | 3.5% | 0.6% | | | |
| 5 | 5.8% | 7.1% | 0.6% | 4.3% | 1.3% | 2.4% | 4.1% | 10.7% | 2.0% | 3.5% | 0.8% | | | | |
| 6 | 3.2% | 0.5% | 3.9% | 0.7% | 3.2% | 7.0% | 13.4% | 1.3% | 5.0% | 1.2% | | | | | |
| 7 | 0.5% | 2.0% | 1.3% | 2.1% | 10.0% | 17.1% | 1.2% | 2.4% | 0.8% | | | | | | |
| 8 | 2.0% | 1.0% | 2.6% | 7.9% | 31.9% | 2.5% | 2.3% | 1.2% | | | | | | | |
| 9 | 0.6% | 2.9% | 3.9% | 21.4% | 1.7% | 2.8% | 1.4% | | | | | | | | |
| 10 | 4.1% | 5.3% | 17.3% | 1.8% | 3.0% | 1.2% | | | | | | | | | |
| 11 | 2.6% | 3.4% | 2.8% | 1.4% | 3.0% | | | | | | | | | | |
| 12 | 3.9% | 4.3% | 0.7% | 3.5% | | | | | | | | | | | |
| 13 | 2.2% | 0.7% | 2.6% | | | | | | | | | | | | |
| 14 | 0.8% | 6.7% | | | | | | | | | | | | | |
| 15 | 3.7% | | | | | | | | | | | | | | |

Table 6-3 Proportionate contributions of array-specific shock to cell total – Triangle 1, data set 3

| Accident period | Development period | | | | | | | | | | | | | | |
|---|---|---|---|---|---|---|---|---|---|---|---|---|---|---|---|
| | 1 | 2 | 3 | 4 | 5 | 6 | 7 | 8 | 9 | 10 | 11 | 12 | 13 | 14 | 15 |
| 1 | 0.3% | 0.3% | 0.5% | 0.3% | 0.5% | 0.5% | 0.4% | 0.6% | 0.3% | 0.8% | 0.3% | 0.4% | 0.5% | 0.4% | 0.3% |
| 2 | 0.3% | 0.3% | 0.3% | 0.6% | 0.3% | 0.4% | 0.5% | 0.3% | 0.3% | 0.3% | 0.4% | 0.3% | 0.2% | 0.4% | |
| 3 | 0.1% | 0.1% | 0.2% | 0.1% | 0.1% | 0.1% | 0.1% | 0.1% | 0.1% | 0.1% | 0.2% | 0.1% | 0.2% | | |
| 4 | 1.1% | 0.4% | 0.5% | 0.7% | 0.7% | 1.0% | 0.6% | 0.7% | 0.5% | 0.7% | 0.7% | 0.5% | | | |
| 5 | 0.6% | 1.1% | 0.9% | 1.2% | 0.8% | 0.4% | 0.4% | 0.3% | 0.6% | 0.7% | 0.6% | | | | |
| 6 | 0.0% | 0.0% | 0.1% | 0.0% | 0.0% | 0.0% | 0.0% | 0.0% | 0.0% | 0.0% | | | | | |
| 7 | 0.7% | 0.6% | 0.9% | 0.4% | 1.1% | 0.6% | 0.4% | 0.5% | 0.7% | | | | | | |
| 8 | 1.0% | 1.1% | 0.8% | 1.4% | 1.7% | 1.4% | 0.8% | 1.7% | | | | | | | |
| 9 | 1.0% | 1.5% | 1.1% | 1.9% | 1.5% | 1.5% | 3.1% | | | | | | | | |
| 10 | 0.1% | 0.1% | 0.1% | 0.1% | 0.1% | 0.1% | | | | | | | | | |
| 11 | 0.4% | 0.2% | 0.4% | 0.6% | 0.3% | | | | | | | | | | |
| 12 | 0.1% | 0.2% | 0.1% | 0.1% | | | | | | | | | | | |
| 13 | 2.3% | 2.3% | 2.2% | | | | | | | | | | | | |
| 14 | 0.1% | 0.1% | | | | | | | | | | | | | |
| 15 | 0.0% | | | | | | | | | | | | | | |

The tables are heat maps in which colouring toward the red end of the spectrum indicates shock contributions that are larger than the triangle average; and the blue end of the spectrum indicates shock contributions that are smaller.

Due to the random nature of simulation, the proportionate contribution of a shock to a cell total in the simulated data is not constant across all cells. However, it can be observed from Table 6-2, Table 6-3 and Appendix B that these contributions are neither systematically skewed toward cells with small values in high development periods, nor are they overwhelmed in cells with larger values from early development periods, as occurs in unbalanced common shock models.

Split umbrella diagonal shocks are clear in Table 6-2 where diagonal patterns are visible and apply to only the relevant subset of accident periods. Row-wise array specific shocks are also exhibited in Table 6-3. As shown in Appendix B for data set 1 where shocks are cell-wise, no clear patterns are observed for proportionate contributions of umbrella and array-specific shocks provided. On the other hand, row-wise shock patterns are evident for data set 2 which was simulated from a row-wise dependence model.

# 7. Conclusions

Section 5.1 derive auto-balance conditions for Tweedie common shock models. These are found to be relatively simple, depending only on the coefficients of variation of the common shock and idiosyncratic components of observations; specifically, the relations between these CoVs. In the case of auto-balance the means and variances of observations assume particularly simple forms.

The common shock models considered are of quite general form such as row-wise or diagonal-wise, etc., or shocks that affect even quite irregular shapes within a triangle. Section 5.2 extends the range of models further with the inclusion of an indefinite number of shocks within the model. Shocks or a hybrid type, such as intersections of subsets of rows and columns, are also included. Auto-balance conditions are derived for all of these.

## Acknowledgement
In relation to the first author, this research was supported under Australian Research Council's Linkage Projects funding scheme (project number LP130100723). The views expressed herein are those of the authors and are not necessarily those of the supporting body.

## Appendix A

**Proof of Lemma 2.3.** Divide (2.10) by (2.7) to obtain

$$\frac{\theta}{\alpha - 1} = \frac{1}{\mu v}. \tag{A. 1}$$

Substitute (A. 1) into (2.10) to obtain

$$\lambda = \frac{1}{v}(\mu v)^\alpha = \mu^\alpha v^{\alpha - 1}. \tag{A. 2}$$

The lemma follows from (A. 1) and (A. 2). ∎

**Proof of Lemma 3.1.** Consider the last two summands of (2.19), i.e. $\beta_{ij}^{(n)} W_{\pi(i,j)}^{(n)} + Z_{ij}^{(n)}$ for $p \neq 1$. By (2.16),

$$\beta_{ij}^{(n)} W_{\pi(i,j)}^{(n)} + Z_{ij}^{(n)} \sim Tw_p^*\left(\theta_{ij}^{(n)}, \left(\frac{\theta_{\pi(i,j)}^{(n)}}{\theta_{ij}^{(n)}}\right)^\alpha \lambda_{\pi(i,j)}^{(n)} + \lambda_{ij}^{(n)}\right),$$

provided that $\beta_{ij}^{(n)} = \theta_{\pi(i,j)}^{(n)} / \theta_{ij}^{(n)}$.

Similarly,

$$X_{ij}^{(n)} = \alpha_{ij}^{(n)} W_{\pi(i,j)}$$
$$+ \left(\beta_{ij}^{(n)} W_{\pi(i,j)}^{(n)} + Z_{ij}^{(n)}\right) \sim Tw_p^*\left(\theta_{ij}^{(n)}, \left(\frac{\theta_{\pi(i,j)}^{(n)}}{\theta_{ij}^{(n)}}\right)^\alpha \lambda_{\pi(i,j)}\right.$$
$$\left. + \left(\frac{\theta_{\pi(i,j)}^{(n)}}{\theta_{ij}^{(n)}}\right)^\alpha \lambda_{\pi(i,j)}^{(n)} + \lambda_{ij}^{(n)}\right), \tag{A. 3}$$

provided that $\alpha_{ij}^{(n)} = \theta_{\pi(i,j)}^{(n)} / \theta_{ij}^{(n)}$.

The case $p = 1$ is similarly treated. This proves the sufficiency of the conditions on $\alpha_{ij}^{(n)} \beta_{ij}^{(n)}$. For necessity, consider the cgf of $X_{ij}^{(n)}$ in the case $p \neq 0,1,2$. It is

$$K_{X_{ij}^{(n)}}(t) = K_{W_{\pi(i,j)}}\left(\alpha_{ij}^{(n)} t\right) + K_{W_{\pi(i,j)}^{(n)}}\left(\beta_{ij}^{(n)} t\right) + K_{Z_{ij}^{(n)}}(t). \tag{A. 4}$$

Consider the first summand on the right:

$$K_{W_{\pi(i,j)}}\left(\alpha_{ij}^{(n)}t\right) = \lambda_{\pi(i,j)}\left[b_p\left(\theta_{\pi(i,j)} + \alpha_{ij}^{(n)}t\right) - b_p\left(\theta_{\pi(i,j)}\right)\right]$$

$$= \lambda_{\pi(i,j)} b_p\left(\theta_{\pi(i,j)}\right)\left[\left(1 + \frac{\alpha_{ij}^{(n)}}{\theta_{\pi(i,j)}}t\right)^{\alpha} - 1\right]. \tag{A.5}$$

by (2.2) and (2.6).

Similarly for the other two summands in (A. 4), whereupon

$$K_{X_{ij}^{(n)}}(t) = b_p\left(\theta_{\pi(i,j)}\right)\left\{\xi_1\left[\left(1 + \frac{\alpha_{ij}^{(n)}}{\theta_{\pi(i,j)}}t\right)^{\alpha} - 1\right] + \xi_2\left[\left(1 + \frac{\beta_{ij}^{(n)}}{\theta_{\pi(i,j)}^{(n)}}t\right)^{\alpha} - 1\right]\right.$$

$$\left. + \xi_3\left[\left(1 + \frac{1}{\theta_{ij}^{(n)}}t\right)^{\alpha} - 1\right]\right\}, \tag{A.6}$$

for constants $\xi_k > 0$, $k = 1,2,3$, and, if this is be $Tw_p^*$, it must be equal to

$$K_{X_{ij}^{(n)}}(t) = b_p\left(\theta_{\pi(i,j)}\right)\xi[(1 + \eta t)^{\alpha} - 1], \tag{A.7}$$

for some constants $\xi, \eta > 0$.

Note that, for $p \neq 0,1,2$, one finds $\alpha \in (-\infty, 0) \cup (0,1)$. For identity between (A. 6) and (A. 7) to hold, all three of the terms raised to power $\alpha$ in (A. 6) must be equal, yielding

$$\frac{\alpha_{ij}^{(n)}}{\theta_{\pi(i,j)}} = \frac{\beta_{ij}^{(n)}}{\theta_{\pi(i,j)}^{(n)}} = \frac{1}{\theta_{ij}^{(n)}}.$$

This proves the necessity of the of the conditions on $\alpha_{ij}^{(n)} \beta_{ij}^{(n)}$ (as well as re-establishing their sufficiency).

The cases $p = 0,1,2$ respectively may be dealt with by a similar analysis of cgf $K_{X_{ij}^{(n)}}(t)$. ∎

**Proof of Lemma 4.1.** For $p \neq 1$, the value of $E\left[X_{ij}^{(n)}\right]$ is obtained from (2.7) with $\theta, \lambda$ replaced by the two arguments on the right side of (3.5). This immediately yields (4.1).

For $p = 1$, by (2.11) and (3.1),

$$E\left[X_{ij}^{(n)}\right] = \left(\lambda_{\pi(i,j)} + \lambda_{\pi(i,j)}^{(n)} + \lambda_{ij}^{(n)}\right) \exp \theta_{ij}^{(n)} = \lambda_{ij}^{(n)} \exp \theta_{ij}^{(n)}\left(\frac{\lambda_{\pi(i,j)}}{\lambda_{ij}^{(n)}} + \frac{\lambda_{\pi(i,j)}^{(n)}}{\lambda_{ij}^{(n)}} + 1\right),$$

which is equal to (4.1) when (2.11) is used again, and it is recalled from Lemma 3.1 that $\theta_{\pi(i,j)} = \theta_{\pi(i,j)}^{(n)} = \theta_{ij}^{(n)}$.

Evaluation of $Var\left[X_{ij}^{(n)}\right]$ for $p \neq 1$ follows the same logic but with (2.7) replaced by (2.8), leading to

$$Var\left[X_{ij}^{(n)}\right] = \left(\mu_{ij}^{(n)}\right)^2 v_{ij}^{(n)} \left(\frac{v_{ij}^{(n)}}{v_{\pi(i,j)}} + \frac{v_{ij}^{(n)}}{v_{\pi(i,j)}^{(n)}} + 1\right),$$

which is identical to (4.2). ■

**Proof of Proposition 5.1.** As stated in the preamble to the proposition, a necessary and sufficient condition for balance is that, for each $n$,

$$\frac{v_{ij}^{(n)}}{v_{\pi(i,j)}^{(n)}} \text{ must be independent of } i,j; \text{ and} \tag{A.8}$$

$$\frac{v_{ij}^{(n)}}{v_{\pi(i,j)}} \text{ must be independent of } i,j. \tag{A.9}$$

These are exactly conditions (a) and (b).

Substitution of conditions (a) and (b) of the proposition in (4.1) and (4.2) yields (5.1)-(5.3) immediately. ■

**Proof of Proposition 5.8.** Commence with condition (a) of Proposition 5.7 for fixed $i,j,n$ and $r = r_1$, and note that $v_{\pi_{[r_1]}(i,j)}$ is constant over all cells $(k,\ell) \in \pi_{[r_1]}(i,j)$. This implies that $v_{k\ell}^{(n)}$ is constant over the same set.

Now, choose any fixed but arbitrary $r = r^*$. Either there is a partition subset $\pi_{[r^*]}(k,\ell)$ that lies within the equivalence class $\mathcal{E}(i,j)$, or there is not. If not, then all subsets of partition $\mathcal{P}_{[r^*]}^{(n)}$ are disconnected from $(i,j)$.

Consider just the former (connected) case, and let cell $(k,\ell)$ be arbitrary within the subset. The cells $(i,j)$ and $(k,\ell)$ are connected and so, by definition, there exists a sequence of subsets $\left\{\mathcal{P}_{[r_g]s_g}^{(n)}, g = 1, \ldots, G\right\}$ such that $(i,j) \in \mathcal{P}_{[r_1]s_1}^{(n)} = \pi_{[r_1]}(i,j), (k,\ell) \in \mathcal{P}_{[r_G]s_G}^{(n)} = \pi_{[r^*]}(k,\ell)$ and $\mathcal{P}_{[r_g]s_g}^{(n)} \cap \mathcal{P}_{[r_{g+1}]s_{g+1}}^{(n)} \neq \emptyset, g = 1, \ldots, G-1$.

Consider the case $g = 1$, for which $\pi_{[r_1]}(i,j)$ and $\mathcal{P}_{[r_2]s_2}^{(n)}$ intersect. It follows that $\mathcal{P}_{[r_2]s_2}^{(n)}$ contains at least one cell $(k_2, \ell_2)$ which lies within $\pi_{[r_1]}(i,j)$ so that $v_{k_2\ell_2}^{(n)} = C^{(n)} v_{\pi_{[r_1]}(i,j)}$, by condition (a) of Proposition 5.7. But recall that $\mathcal{P}_{[r_2]s_2}^{(n)}$ is just a partition subset, and so can be represented as $\pi_{[r_2]}(k_2, \ell_2)$, and so, by the same condition, $v_{k_2\ell_2}^{(n)} = C^{(n)} v_{\pi_{[r_2]}(k_2, \ell_2)}$. From this, it follows that $v_{\pi_{[r_2]}(k_2, \ell_2)} = v_{\pi_{[r_1]}(i,j)}$.

The argument can be extended along the chain of subsets connecting $(i,j)$ to $(k,\ell)$, yielding the result that $v_{\pi_{[r_{g+1}]}(k_{g+1}, \ell_{g+1})} = v_{\pi_{[r_g]}(k_g, \ell_g)}$ for $g = 1, \ldots, G-1$, where $\{(k_g, \ell_g), g = 1, \ldots, G\}$ is a connected sequence of cells with $(k_1, \ell_1) = (i,j)$ and $(k_G, \ell_G) = (k, \ell)$. One may now conclude that $v_{\pi_{[r_1]}(i,j)} = v_{\pi_{[r_1]}(k_1, \ell_1)} = v_{\pi_{[r_G]}(k_G, \ell_G)} = v_{\pi_{[r^*]}(k, \ell)}$.

Recall that $r_1, r^*$ were chosen arbitrarily; for given $r^*$, $\pi_{[r^*]}(k,\ell)$ was chosen arbitrarily subject to connectedness to $(i,j)$; and $(k,\ell)$ is an arbitrary cell in $\pi_{[r^*]}(k,\ell)$. It follows that, for any choice of $r_1, r^*$, $v_{\pi_{[r_1]}(i,j)} = v_{\pi_{[r^*]}(k,\ell)}$ for any cell $(k,\ell)$ that is connected to $(i,j)$. This proves the proposition in respect of $v_{\pi_{[r]}(k,\ell)}$. The proof in respect of $v^{(n)}_{\pi_{[r]}(k,\ell)}$ is entirely parallel. ∎

# Appendix B

# Parameter specifications of synthetic data

| | Data set 1 (cell-wise dependence) | Data set 2 (row-wise dependence) | Data set 3 (split diagonal umbrella shock and row-wise array-specific shock) |
|---|---|---|---|
| **Power parameter** | | p = 1.8 | |
| **Umbrella shock multiple** | $C^{(1)} = 0.6^4$ <br> $C^{(2)} = 0.3^4$ | $C^{(1)} = 0.44^4$ <br> $C^{(2)} = 0.3^4$ | $C^{(1)} = 0.45^4$ <br> $C^{(2)} = 0.38^4$ |
| **Array-specific shock multiple** | $K^{(1)} = 0.33^4$ <br> $K^{(2)} = 0.45^4$ | $K^{(1)} = 0.29^4$ <br> $K^{(2)} = 0.45^4$ | $K^{(1)} = 0.31^4$ <br> $K^{(2)} = 0.5^4$ |
| **Umbrella shock – Expected value** | $\mu_{\pi(i,j)} = \{100, 500, 1000, 1000, 1000, 1000, 500, 250, 100, 50, 50, 50, 50, 50, 50\}$ for j = 1,...,15 respectively of partition subsets $\{\{X_{1j}^{(n)}\}: \text{all } n\}$. <br><br> $\mu_{\pi(i,j)} = 1.02\mu_{\pi(i-1,j)}$ for partition subsets $\{\{X_{ij}^{(n)}\}: i > 1, \text{all } j, \text{all } n\}$. | $\mu_{\pi(i,j)} = 100$ for partition subsets $\{\mathcal{R}_1^{(n)}: \text{all } n\}$. <br><br> $\mu_{\pi(i,j)} = 1.02\mu_{\pi(i-1,j)}$ for partition subsets $\{\mathcal{R}_i^{(n)}: i > 1, \text{all } n\}$. | $\mu_{\pi(i,t-i+1)} = \{100, 102, 104, 106, 108, 110, 113, 115, 117, 120, 117, 119, 121, 124, 126\}$ for t = 1, ...,15 respectively of partition subsets $\{\mathcal{D}_t^{(n)}: \text{all } n, i \leq 10\}$. <br><br> $\mu_{\pi(i,t-i+1)} = \{146, 149, 152, 153, 158\}$ for t = 11,...,15 respectively of partition subsets $\{\mathcal{D}_t^{(n)}: \text{all } n, i > 10\}$. |
| **Umbrella shock - CoV** | $\sqrt{v_{\pi(i,j)}} = \frac{1}{0.6^2} * \{0.1, 0.06, 0.05, 0.05, 0.06, 0.06, 0.1, 0.15, 0.2, 0.3, 0.45, 0.6, 0.75, 0.9, 0.9\}$ for j = 1,...,15 respectively of partition subsets $\{\{X_{ij}^{(n)}\}: \text{all } n, \text{all } i\}$. | $\sqrt{v_{\pi(i,j)}} = \frac{1}{0.44^2} * \{0.1, 0.09, 0.08, 0.09, 0.1, 0.1, 0.11, 0.09, 0.12, 0.1, 0.08, 0.09, 0.08, 0.08\}$ for j = 1,...,15 respectively of partition subsets $\{\mathcal{R}_i^{(n)}: \text{all } n\}$. | $\sqrt{v_{\pi(i,j)}} = \frac{0.1}{0.45^2}$ for partition subsets $\{\mathcal{R}_i^{(n)} \cup \mathcal{D}_t^{(n)}: \text{all } n, \text{all } t, i \leq 10\}$, or in short, all cells with i ≤ 10. <br><br> $\sqrt{v_{\pi(i,j)}} = \frac{0.08}{0.45^2}$ for partition subsets |
| | | | $\left\{\mathcal{R}_i^{(n)} \cup \mathcal{D}_t^{(n)}: \text{all n, all t, i} > 10\right\}$, or in short, all cells with i > 10. |
| **Array-specific shock - Expected value** | $\mu_{\pi(i,j)}^{(1)} = \{25, 50, 100, 100, 100, 25, 20,$ $20, 10, 10, 10, 10, 10, 10, 10\}$ for j = $1, \ldots, 15$ respectively of partition subsets $\left\{X_{1j}^{(n)}\right\}$. $\mu_{\pi(i,j)}^{(1)} = 1.02 \mu_{\pi(i-1,j)}^{(1)}$ for partition subsets $\left\{\left\{X_{ij}^{(n)}\right\}: i > 1, \text{all j}\right\}$. $\mu_{\pi(i,j)}^{(2)} = \{1000, 1000, 1000, 500, 200,$ $20, 10, 5, 5, 5, 5, 5, 5, 5, 5\}$ for j = $1, \ldots, 15$ *respectively of* partition subsets $\left\{\left\{X_{ij}^{(2)}\right\}: \text{all i}\right\}$. | $\mu_{\pi(i,j)}^{(1)} = 25$ for partition subset $\mathcal{R}_1^{(n)}$. $\mu_{\pi(i,j)}^{(1)} = 1.02 \mu_{\pi(i-1,j)}^{(1)}$ for partition subsets $\left\{\mathcal{R}_i^{(1)}: i > 1,\right\}$. $\mu_{\pi(i,j)}^{(2)} = 1000$ for partition subsets $\left\{\mathcal{R}_i^{(2)}: \text{all i}\right\}$. | |
| **Array-specific shock - CoV** | $\sqrt{v_{\pi(i,j)}^{(n)}} = \sqrt{\frac{C^{(n)}}{K^{(n)}}} \sqrt{v_{\pi(i,j)}}$ for all partition subsets. | | |
| **Idiosyncratic component - Expected value** | $\mu_{1j}^{(1)} = \{500, 1000, 1500, 2000, 2000, 1000, 700, 500, 400, 200, 100, 50, 25, 15, 10\}$ *for j* $= 1, \ldots, 15$ *respectively*. $\mu_{ij}^{(1)} = 1.02 \mu_{i-1,j}^{(1)}$ *for i* $= 2, \ldots, 15,$ *and all j*. $\mu_{ij}^{(2)} = \{3000, 4000, 1000, 500, 400, 300, 200, 100, 100, 100, 100, 50, 50, 50, 50\}$ *for j* $= 1, \ldots 15$ *respectively and all i*. | | |
| **Idiosyncratic component - CoV** | $\sqrt{v_{ij}^{(n)}} = \sqrt{\frac{1}{K^{(n)}}} \sqrt{v_{\pi(i,j)}^{(n)}}$ for all partition subsets. | | |

**Data set 1 (cell-wise dependence) - Proportionate contributions of shocks to cell total: triangle 1, umbrella shock (top left); triangle 1, array-specific shock (top right); triangle 2, umbrella shock (bottom left); triangle 2, array-specific shock (bottom right)**

| Accident period | Development period | | | | | | | | | | | | | | |
|---|---|---|---|---|---|---|---|---|---|---|---|---|---|---|---|
| | 1 | 2 | 3 | 4 | 5 | 6 | 7 | 8 | 9 | 10 | 11 | 12 | 13 | 14 | 15 |
| 1 | 13.5% | 13.1% | 6.3% | 14.3% | 7.2% | 8.8% | 0.1% | 10.8% | 11.2% | 35.1% | 75.6% | 1.0% | 13.1% | 4.2% | 25.6% |
| 2 | 5.1% | 12.5% | 16.6% | 28.2% | 3.3% | 6.7% | 13.8% | 5.5% | 4.5% | 27.5% | 1.9% | 3.2% | 1.4% | 32.3% | |
| 3 | 21.1% | 6.9% | 8.9% | 8.5% | 10.2% | 17.4% | 6.8% | 15.1% | 11.5% | 15.7% | 59.0% | 8.0% | 4.3% | | |
| 4 | 13.4% | 10.0% | 7.1% | 10.1% | 4.9% | 3.8% | 11.3% | 15.4% | 19.2% | 2.3% | 55.5% | 95.7% | | | |
| 5 | 6.5% | 15.2% | 8.4% | 15.9% | 9.3% | 16.3% | 9.1% | 15.9% | 8.0% | 7.5% | 3.2% | | | | |
| 6 | 4.0% | 6.4% | 13.0% | 21.0% | 14.4% | 10.2% | 7.6% | 17.7% | 8.9% | 77.4% | | | | | |
| 7 | 6.2% | 9.5% | 12.5% | 13.0% | 12.4% | 16.6% | 11.5% | 3.8% | 23.0% | | | | | | |
| 8 | 19.4% | 11.0% | 16.0% | 16.9% | 10.8% | 11.1% | 6.1% | 7.9% | | | | | | | |
| 9 | 6.4% | 11.8% | 9.1% | 6.6% | 6.2% | 15.8% | 26.9% | | | | | | | | |
| 10 | 18.9% | 13.0% | 17.5% | 6.8% | 9.3% | 10.6% | | | | | | | | | |
| 11 | 6.5% | 6.0% | 13.0% | 9.2% | 7.1% | | | | | | | | | | |
| 12 | 4.3% | 7.7% | 11.3% | 26.4% | | | | | | | | | | | |
| 13 | 16.9% | 19.0% | 9.9% | | | | | | | | | | | | |
| 14 | 21.3% | 16.8% | | | | | | | | | | | | | |
| 15 | 28.9% | | | | | | | | | | | | | | |

| Accident period | Development period | | | | | | | | | | | | | | |
|---|---|---|---|---|---|---|---|---|---|---|---|---|---|---|---|
| | 1 | 2 | 3 | 4 | 5 | 6 | 7 | 8 | 9 | 10 | 11 | 12 | 13 | 14 | 15 |
| 1 | 1.8% | 0.3% | 0.5% | 1.5% | 0.2% | 0.5% | 0.6% | 0.7% | 1.3% | 0.0% | 0.0% | 1.1% | 0.0% | 0.0% | 8.3% |
| 2 | 1.2% | 0.6% | 1.1% | 1.8% | 1.1% | 0.3% | 0.4% | 0.0% | 0.0% | 0.9% | 0.0% | 0.0% | 0.0% | 0.0% | |
| 3 | 0.1% | 1.4% | 0.3% | 0.1% | 0.6% | 0.4% | 0.7% | 0.1% | 0.1% | 0.0% | 0.0% | 0.7% | 21.5% | | |
| 4 | 0.5% | 1.7% | 0.6% | 0.9% | 0.1% | 2.2% | 0.1% | 4.5% | 0.0% | 2.2% | 0.1% | 0.0% | | | |
| 5 | 0.0% | 3.8% | 1.5% | 2.5% | 0.8% | 0.4% | 3.1% | 0.1% | 0.0% | 0.0% | 0.1% | | | | |
| 6 | 0.1% | 0.3% | 1.1% | 1.3% | 0.8% | 1.9% | 0.1% | 3.2% | 0.1% | 0.0% | | | | | |
| 7 | 0.0% | 1.4% | 1.0% | 0.5% | 2.3% | 1.1% | 0.3% | 0.1% | 0.2% | | | | | | |
| 8 | 0.1% | 3.4% | 1.0% | 0.7% | 0.4% | 0.3% | 0.1% | 2.2% | | | | | | | |
| 9 | 5.7% | 1.7% | 2.6% | 1.4% | 2.9% | 1.2% | 0.5% | | | | | | | | |
| 10 | 0.6% | 0.1% | 2.4% | 0.1% | 0.6% | 0.7% | | | | | | | | | |
| 11 | 0.9% | 0.7% | 1.1% | 0.3% | 0.5% | | | | | | | | | | |
| 12 | 1.0% | 0.7% | 1.4% | 1.0% | | | | | | | | | | | |
| 13 | 1.3% | 0.2% | 1.2% | | | | | | | | | | | | |
| 14 | 1.9% | 1.7% | | | | | | | | | | | | | |
| 15 | 0.0% | | | | | | | | | | | | | | |

**Triangle 1, umbrella shock (top left):**

| Accident period | \multicolumn{15}{c}{Development period} |
|---|---|---|---|---|---|---|---|---|---|---|---|---|---|---|---|
| | 1 | 2 | 3 | 4 | 5 | 6 | 7 | 8 | 9 | 10 | 11 | 12 | 13 | 14 | 15 |
| 1 | 1.2% | 0.8% | 0.5% | 0.6% | 0.5% | 0.6% | 0.0% | 0.9% | 0.6% | 1.6% | 4.4% | 0.0% | 0.9% | 1.2% | 1.7% |
| 2 | 0.3% | 0.8% | 0.9% | 1.3% | 0.2% | 0.6% | 0.8% | 0.4% | 0.9% | 1.4% | 0.2% | 0.1% | 0.1% | 0.3% | |
| 3 | 1.4% | 0.4% | 0.8% | 0.6% | 0.7% | 1.6% | 0.6% | 0.8% | 1.0% | 0.9% | 3.0% | 0.3% | 0.2% | | |
| 4 | 1.0% | 0.7% | 0.4% | 0.8% | 0.4% | 0.4% | 0.7% | 0.7% | 2.9% | 0.2% | 6.0% | 0.4% | | | |
| 5 | 0.7% | 0.6% | 0.4% | 1.2% | 0.5% | 1.0% | 0.5% | 1.4% | 0.4% | 0.3% | 0.4% | | | | |
| 6 | 0.4% | 0.4% | 0.8% | 0.8% | 0.8% | 0.6% | 0.9% | 1.0% | 1.0% | 4.5% | | | | | |
| 7 | 0.7% | 0.4% | 0.8% | 1.1% | 0.9% | 1.4% | 1.4% | 0.2% | 1.2% | | | | | | |
| 8 | 1.3% | 0.5% | 0.9% | 1.3% | 1.0% | 0.9% | 0.7% | 0.5% | | | | | | | |
| 9 | 0.3% | 0.4% | 0.4% | 0.6% | 0.5% | 1.0% | 2.1% | | | | | | | | |
| 10 | 1.9% | 1.7% | 0.8% | 0.4% | 0.8% | 0.6% | | | | | | | | | |
| 11 | 0.5% | 0.3% | 0.6% | 0.8% | 0.5% | | | | | | | | | | |
| 12 | 0.3% | 1.0% | 0.8% | 1.7% | | | | | | | | | | | |
| 13 | 0.9% | 0.9% | 0.5% | | | | | | | | | | | | |
| 14 | 1.6% | 0.8% | | | | | | | | | | | | | |
| 15 | 1.2% | | | | | | | | | | | | | | |

**Triangle 1, array-specific shock (top right):**

| Accident period | \multicolumn{15}{c}{Development period} |
|---|---|---|---|---|---|---|---|---|---|---|---|---|---|---|---|
| | 1 | 2 | 3 | 4 | 5 | 6 | 7 | 8 | 9 | 10 | 11 | 12 | 13 | 14 | 15 |
| 1 | 4.4% | 2.5% | 3.7% | 4.1% | 5.2% | 2.7% | 3.8% | 1.4% | 3.6% | 4.6% | 7.1% | 1.5% | 6.4% | 0.0% | 5.2% |
| 2 | 3.7% | 4.7% | 3.8% | 2.9% | 3.2% | 1.6% | 2.8% | 5.2% | 2.5% | 11.9% | 5.2% | 3.2% | 2.2% | 4.5% | |
| 3 | 3.6% | 4.9% | 5.3% | 3.5% | 3.5% | 4.6% | 2.0% | 4.0% | 5.0% | 2.4% | 7.1% | 3.8% | 1.6% | | |
| 4 | 4.8% | 3.2% | 3.8% | 3.3% | 6.9% | 4.9% | 4.2% | 2.4% | 3.8% | 1.4% | 3.3% | 2.2% | | | |
| 5 | 2.8% | 3.0% | 3.2% | 3.2% | 2.9% | 2.9% | 5.8% | 5.5% | 3.1% | 1.9% | 3.3% | | | | |
| 6 | 2.9% | 3.2% | 4.5% | 3.4% | 6.4% | 3.5% | 9.0% | 3.7% | 7.5% | 3.6% | | | | | |
| 7 | 2.1% | 5.0% | 4.8% | 4.4% | 4.4% | 3.6% | 1.1% | 2.3% | 7.6% | | | | | | |
| 8 | 2.8% | 3.7% | 2.4% | 4.4% | 4.6% | 4.8% | 1.1% | 3.6% | | | | | | | |
| 9 | 4.0% | 2.4% | 4.2% | 3.2% | 4.0% | 2.3% | 4.4% | | | | | | | | |
| 10 | 4.7% | 4.6% | 4.4% | 4.7% | 4.7% | 3.9% | | | | | | | | | |
| 11 | 4.1% | 1.6% | 3.9% | 5.5% | 4.7% | | | | | | | | | | |
| 12 | 3.0% | 3.5% | 4.2% | 3.7% | | | | | | | | | | | |
| 13 | 4.1% | 2.7% | 4.5% | | | | | | | | | | | | |
| 14 | 2.7% | 4.1% | | | | | | | | | | | | | |
| 15 | 2.7% | | | | | | | | | | | | | | |

**Data set 2 (row-wise dependence) - Proportionate contributions of shocks to cell total: triangle 1, umbrella shock (top left); triangle 1, array-specific shock (top right); triangle 2, umbrella shock (bottom left); triangle 2, array-specific shock (bottom right)**

**Triangle 2, umbrella shock (bottom left):**

| Accident period | \multicolumn{15}{c}{Development period} |
|---|---|---|---|---|---|---|---|---|---|---|---|---|---|---|---|
| | 1 | 2 | 3 | 4 | 5 | 6 | 7 | 8 | 9 | 10 | 11 | 12 | 13 | 14 | 15 |
| 1 | 5.4% | 9.5% | 3.8% | 4.0% | 3.8% | 4.4% | 2.1% | 3.2% | 2.1% | 3.0% | 3.5% | 3.0% | 1.9% | 3.1% | 2.7% |
| 2 | 3.5% | 3.5% | 7.3% | 6.6% | 4.5% | 4.2% | 4.7% | 3.2% | 4.5% | 2.6% | 3.1% | 3.7% | 9.2% | 3.7% | |
| 3 | 3.2% | 3.1% | 3.5% | 4.4% | 4.2% | 4.8% | 3.2% | 3.4% | 3.8% | 3.6% | 3.2% | 6.8% | 5.1% | | |
| 4 | 7.7% | 14.3% | 10.5% | 11.6% | 15.9% | 17.3% | 10.9% | 10.4% | 17.7% | 14.1% | 14.0% | 12.4% | | | |
| 5 | 2.5% | 3.9% | 4.0% | 4.4% | 5.3% | 2.2% | 5.9% | 2.6% | 2.0% | 2.5% | 4.0% | | | | |
| 6 | 4.0% | 3.0% | 2.1% | 3.3% | 2.7% | 3.4% | 2.6% | 1.6% | 3.0% | 5.9% | | | | | |
| 7 | 0.4% | 0.5% | 0.5% | 0.7% | 0.4% | 0.9% | 0.6% | 0.7% | 0.4% | | | | | | |
| 8 | 2.7% | 3.1% | 2.5% | 2.1% | 3.3% | 3.1% | 3.3% | 3.5% | | | | | | | |
| 9 | 8.5% | 3.8% | 8.1% | 9.7% | 8.6% | 4.3% | 6.2% | | | | | | | | |
| 10 | 1.5% | 1.6% | 1.9% | 1.3% | 1.9% | 1.8% | | | | | | | | | |
| 11 | 2.1% | 2.2% | 2.2% | 2.3% | 2.0% | | | | | | | | | | |
| 12 | 1.2% | 1.2% | 1.4% | 1.2% | | | | | | | | | | | |
| 13 | 2.4% | 2.3% | 10.3% | | | | | | | | | | | | |
| 14 | 7.1% | 7.9% | | | | | | | | | | | | | |
| 15 | 3.1% | | | | | | | | | | | | | | |

**Triangle 2, array-specific shock (bottom right):**

| Accident period | \multicolumn{15}{c}{Development period} |
|---|---|---|---|---|---|---|---|---|---|---|---|---|---|---|---|
| | 1 | 2 | 3 | 4 | 5 | 6 | 7 | 8 | 9 | 10 | 11 | 12 | 13 | 14 | 15 |
| 1 | 0.6% | 1.1% | 0.4% | 0.5% | 0.4% | 0.5% | 0.2% | 0.4% | 0.2% | 0.3% | 0.4% | 0.4% | 0.2% | 0.4% | 0.3% |
| 2 | 0.7% | 0.7% | 1.5% | 1.4% | 0.9% | 0.9% | 1.0% | 0.7% | 0.9% | 0.5% | 0.6% | 0.8% | 1.9% | 0.8% | |
| 3 | 2.0% | 1.9% | 2.1% | 2.7% | 2.6% | 3.0% | 1.9% | 2.1% | 2.4% | 2.2% | 2.0% | 4.2% | 3.1% | | |
| 4 | 0.3% | 0.6% | 0.4% | 0.5% | 0.7% | 0.7% | 0.4% | 0.4% | 0.7% | 0.6% | 0.6% | 0.5% | | | |
| 5 | 0.0% | 0.0% | 0.0% | 0.0% | 0.0% | 0.0% | 0.0% | 0.0% | 0.0% | 0.0% | 0.0% | | | | |
| 6 | 0.0% | 0.0% | 0.0% | 0.0% | 0.0% | 0.0% | 0.0% | 0.0% | 0.0% | 0.0% | | | | | |
| 7 | 0.9% | 1.0% | 1.1% | 1.4% | 1.0% | 1.9% | 1.2% | 1.6% | 0.8% | | | | | | |
| 8 | 0.1% | 0.1% | 0.1% | 0.1% | 0.1% | 0.1% | 0.1% | 0.1% | | | | | | | |
| 9 | 0.8% | 0.3% | 0.7% | 0.9% | 0.8% | 0.4% | 0.6% | | | | | | | | |
| 10 | 0.7% | 0.7% | 0.8% | 0.6% | 0.9% | 0.8% | | | | | | | | | |
| 11 | 0.4% | 0.4% | 0.4% | 0.4% | 0.3% | | | | | | | | | | |
| 12 | 0.4% | 0.4% | 0.5% | 0.4% | | | | | | | | | | | |
| 13 | 0.1% | 0.1% | 0.5% | | | | | | | | | | | | |
| 14 | 0.9% | 1.0% | | | | | | | | | | | | | |
| 15 | 0.5% | | | | | | | | | | | | | | |

### Triangle 2, umbrella shock

| Accident period | \ Development period 1 | 2 | 3 | 4 | 5 | 6 | 7 | 8 | 9 | 10 | 11 | 12 | 13 | 14 | 15 |
|---|---|---|---|---|---|---|---|---|---|---|---|---|---|---|---|
| 1 | 1.0% | 0.9% | 1.0% | 0.6% | 1.0% | 0.7% | 0.6% | 0.8% | 0.7% | 0.7% | 0.7% | 0.6% | 0.6% | 0.7% | 0.5% |
| 2 | 1.1% | 1.3% | 1.0% | 1.2% | 0.7% | 0.7% | 1.3% | 0.9% | 0.8% | 0.9% | 1.3% | 1.1% | 0.9% | 1.0% | |
| 3 | 0.8% | 1.0% | 0.6% | 0.8% | 1.0% | 1.2% | 0.7% | 1.0% | 0.7% | 0.8% | 0.8% | 1.2% | 1.0% | | |
| 4 | 2.5% | 2.8% | 2.4% | 2.2% | 2.0% | 2.6% | 3.0% | 2.6% | 2.0% | 4.3% | 3.1% | 3.1% | | | |
| 5 | 0.6% | 0.7% | 0.8% | 0.7% | 0.6% | 0.5% | 0.6% | 0.6% | 0.6% | 1.0% | 0.6% | | | | |
| 6 | 0.5% | 0.4% | 0.6% | 0.7% | 0.4% | 0.5% | 0.4% | 0.6% | 0.4% | 0.5% | | | | | |
| 7 | 0.1% | 0.1% | 0.1% | 0.1% | 0.2% | 0.1% | 0.1% | 0.1% | 0.1% | | | | | | |
| 8 | 0.7% | 0.7% | 0.9% | 0.9% | 0.8% | 0.7% | 0.7% | 0.5% | | | | | | | |
| 9 | 1.1% | 1.1% | 1.3% | 1.1% | 1.5% | 1.1% | 1.2% | | | | | | | | |
| 10 | 0.3% | 0.4% | 0.4% | 0.4% | 0.4% | 0.4% | | | | | | | | | |
| 11 | 0.3% | 0.6% | 0.5% | 0.5% | 0.3% | | | | | | | | | | |
| 12 | 0.4% | 0.2% | 0.3% | 0.3% | | | | | | | | | | | |
| 13 | 0.8% | 0.8% | 0.9% | | | | | | | | | | | | |
| 14 | 2.0% | 1.8% | | | | | | | | | | | | | |
| 15 | 0.8% | | | | | | | | | | | | | | |

### Triangle 2, array-specific shock

| Accident period | \ Development period 1 | 2 | 3 | 4 | 5 | 6 | 7 | 8 | 9 | 10 | 11 | 12 | 13 | 14 | 15 |
|---|---|---|---|---|---|---|---|---|---|---|---|---|---|---|---|
| 1 | 4.4% | 3.9% | 4.4% | 2.4% | 4.5% | 2.9% | 2.6% | 3.6% | 2.9% | 3.1% | 3.2% | 2.8% | 2.5% | 2.8% | 2.3% |
| 2 | 2.5% | 3.0% | 2.2% | 2.7% | 1.6% | 1.6% | 2.9% | 2.1% | 1.8% | 2.0% | 2.9% | 2.6% | 2.0% | 2.3% | |
| 3 | 3.3% | 4.2% | 2.6% | 3.4% | 4.5% | 5.3% | 3.3% | 4.5% | 3.2% | 3.7% | 3.6% | 5.4% | 4.6% | | |
| 4 | 7.0% | 7.8% | 6.5% | 6.1% | 5.6% | 7.1% | 8.2% | 7.2% | 5.5% | 11.9% | 8.6% | 8.6% | | | |
| 5 | 3.1% | 3.7% | 4.0% | 3.5% | 3.0% | 2.7% | 2.9% | 2.8% | 2.9% | 5.1% | 3.0% | | | | |
| 6 | 2.6% | 2.0% | 3.0% | 3.2% | 2.0% | 2.4% | 2.0% | 3.1% | 1.7% | 2.4% | | | | | |
| 7 | 0.9% | 1.6% | 1.0% | 1.1% | 2.6% | 0.9% | 1.3% | 1.4% | 1.5% | | | | | | |
| 8 | 2.4% | 2.6% | 3.3% | 3.1% | 3.1% | 2.6% | 2.4% | 1.7% | | | | | | | |
| 9 | 2.0% | 2.0% | 2.4% | 2.0% | 2.6% | 1.9% | 2.1% | | | | | | | | |
| 10 | 3.1% | 4.6% | 4.9% | 4.9% | 4.8% | 4.2% | | | | | | | | | |
| 11 | 5.2% | 9.3% | 8.2% | 7.5% | 5.2% | | | | | | | | | | |
| 12 | 5.9% | 3.9% | 4.3% | 4.5% | | | | | | | | | | | |
| 13 | 2.5% | 2.5% | 2.8% | | | | | | | | | | | | |
| 14 | 5.2% | 4.5% | | | | | | | | | | | | | |
| 15 | 2.7% | | | | | | | | | | | | | | |

**Data set 3 (exotic case) - Proportionate contributions of shocks to cell total: triangle 2, umbrella shock (left); triangle 2, array-specific shock (right)**

### Triangle 2, umbrella shock (exotic)

| Accident period | \ Development period 1 | 2 | 3 | 4 | 5 | 6 | 7 | 8 | 9 | 10 | 11 | 12 | 13 | 14 | 15 |
|---|---|---|---|---|---|---|---|---|---|---|---|---|---|---|---|
| 1 | 0.4% | 2.4% | 1.0% | 1.0% | 2.8% | 3.6% | 0.3% | 0.8% | 0.6% | 1.6% | 2.3% | 8.5% | 1.0% | 1.5% | 0.4% |
| 2 | 3.7% | 1.2% | 1.4% | 4.2% | 2.0% | 0.2% | 1.3% | 0.4% | 1.8% | 2.9% | 14.6% | 0.9% | 1.3% | 0.6% | |
| 3 | 1.4% | 2.2% | 3.6% | 1.6% | 0.3% | 0.8% | 0.4% | 1.8% | 2.4% | 7.5% | 1.0% | 2.1% | 0.5% | | |
| 4 | 1.7% | 2.2% | 1.5% | 0.3% | 0.8% | 0.5% | 2.5% | 2.5% | 8.4% | 0.9% | 2.0% | 0.4% | | | |
| 5 | 6.3% | 2.5% | 0.3% | 1.1% | 0.6% | 1.1% | 2.3% | 9.9% | 0.7% | 2.7% | 0.4% | | | | |
| 6 | 3.1% | 0.1% | 1.2% | 0.5% | 1.7% | 3.5% | 11.8% | 1.2% | 1.7% | 0.5% | | | | | |
| 7 | 0.3% | 0.6% | 0.7% | 1.4% | 4.3% | 12.5% | 1.0% | 1.7% | 0.5% | | | | | | |
| 8 | 0.7% | 0.5% | 1.5% | 3.2% | 9.9% | 0.9% | 1.8% | 0.5% | | | | | | | |
| 9 | 0.5% | 2.0% | 4.7% | 10.2% | 0.8% | 1.7% | 0.4% | | | | | | | | |
| 10 | 1.3% | 3.7% | 8.4% | 1.0% | 2.0% | 0.4% | | | | | | | | | |
| 11 | 0.9% | 2.1% | 0.9% | 0.4% | 2.0% | | | | | | | | | | |
| 12 | 2.8% | 1.4% | 0.4% | 1.1% | | | | | | | | | | | |
| 13 | 1.0% | 0.3% | 1.6% | | | | | | | | | | | | |
| 14 | 0.5% | 1.2% | | | | | | | | | | | | | |
| 15 | 1.0% | | | | | | | | | | | | | | |

### Triangle 2, array-specific shock (exotic)

| Accident period | \ Development period 1 | 2 | 3 | 4 | 5 | 6 | 7 | 8 | 9 | 10 | 11 | 12 | 13 | 14 | 15 |
|---|---|---|---|---|---|---|---|---|---|---|---|---|---|---|---|
| 1 | 2.1% | 1.7% | 1.4% | 1.7% | 2.0% | 4.0% | 2.6% | 1.6% | 2.5% | 2.0% | 1.6% | 1.9% | 2.2% | 2.0% | 2.0% |
| 2 | 6.8% | 4.4% | 6.3% | 7.9% | 6.0% | 6.6% | 7.2% | 4.4% | 5.8% | 5.5% | 8.5% | 5.1% | 4.4% | 8.6% | |
| 3 | 3.8% | 7.8% | 5.3% | 3.7% | 5.3% | 3.3% | 3.4% | 4.5% | 3.5% | 3.4% | 4.5% | 5.9% | 5.9% | | |
| 4 | 4.9% | 2.7% | 2.9% | 5.6% | 2.9% | 3.6% | 5.5% | 3.1% | 3.2% | 3.5% | 4.6% | 4.1% | | | |
| 5 | 8.6% | 5.5% | 4.9% | 4.2% | 5.2% | 2.5% | 3.2% | 4.2% | 3.1% | 6.9% | 4.4% | | | | |
| 6 | 8.8% | 3.4% | 6.1% | 5.3% | 5.2% | 6.2% | 6.6% | 6.6% | 5.7% | 6.2% | | | | | |
| 7 | 6.2% | 2.7% | 7.1% | 4.3% | 7.4% | 6.6% | 5.4% | 5.3% | 6.8% | | | | | | |
| 8 | 3.3% | 4.5% | 4.1% | 5.2% | 4.9% | 4.3% | 5.5% | 5.9% | | | | | | | |
| 9 | 2.4% | 2.9% | 3.9% | 2.6% | 2.0% | 2.7% | 2.5% | | | | | | | | |
| 10 | 3.9% | 6.1% | 4.3% | 5.2% | 6.3% | 5.2% | | | | | | | | | |
| 11 | 4.3% | 3.8% | 3.5% | 5.4% | 6.0% | | | | | | | | | | |
| 12 | 11.9% | 13.2% | 11.2% | 7.9% | | | | | | | | | | | |
| 13 | 2.0% | 2.0% | 2.4% | | | | | | | | | | | | |
| 14 | 12.2% | 7.1% | | | | | | | | | | | | | |
| 15 | 3.0% | | | | | | | | | | | | | | |